\documentclass[9pt,twocolumn]{extarticle}
\pdfoutput=1
\usepackage{soul}
\usepackage{titlesec} 
\usepackage[superscript, nomove]{cite}

\usepackage{float}
\usepackage{caption}
\usepackage{lipsum,ulem}
\usepackage[verbose]{placeins}

\usepackage{dsfont}
\usepackage{graphicx}
\usepackage{subcaption}
\usepackage[margin=0.9in]{geometry}
\usepackage[usenames,dvipsnames]{xcolor}
\usepackage[colorlinks,linkcolor=Blue,urlcolor=Blue,citecolor=Blue]{hyperref}
\usepackage{amsmath,amssymb}
\usepackage[squaren]{SIunits}

\usepackage{mathpazo}
\usepackage{courier}
\normalfont
\usepackage[T1]{fontenc}
\usepackage{caption}
\usepackage{upgreek}

\newcommand{\ad}[1]{\textsuperscript{#1}\kern-2pt}

\makeatletter
\def\blx@maxline{77}
\makeatother

\usepackage{capt-of}

\def\mytitle{Hilbert-space selected switch of helical edges in an artificial quantum Hall insulator}

\title{\vspace{-1.0cm}\huge\textbf{\textrm{\mytitle}}}  

\author{Naijie Ren,$^{1,2*}$ Zhiren Xiong,$^{1,2*}$ Kaining Yang,$^{3*}$ Yanran Shi,$^{4*}$ Hanwen Wang,$^{3}$ Kenji Watanabe,$^{5}$ Takashi Taniguchi,$^{6}$ \\Neng Wan,$^{7,8\dagger}$ Xiaojun Jia,$^{1,2\dagger}$ Jianpeng Liu,$^{3,4\dagger}$ Zheng Vitto Han,$^{1,2,3\dagger}$ Yaning Wang$^{3\dagger}$}

\date{} 
\begin{document}
\twocolumn[{
\maketitle 
\vspace{-5mm}
\begin{center}
\begin{minipage}{1\textwidth}
\begin{center}
\textit{
\\\textsuperscript{1} State Key Laboratory of Quantum Optics and Quantum Optics Devices, Institute of Optoelectronics, Shanxi University, Taiyuan 030006, China
\\\textsuperscript{2} Collaborative Innovation Center of Extreme Optics, Shanxi University, Taiyuan 030006, China
\\\textsuperscript{3}Liaoning Academy of Materials, Shenyang 110167, China
\\\textsuperscript{4}School of Physical Science and Technology, ShanghaiTech Laboratory for Topological Physics,  ShanghaiTech University, Shanghai 201210, China
\\\textsuperscript{5}Research Center for Electronic and Optical Materials, National Institute for Materials Science, 1-1 Namiki, Tsukuba 305-0044, Japan
\\\textsuperscript{6}Research Center for Materials Nanoarchitectonics, National Institute for Materials Science, 1-1 Namiki, Tsukuba 305-0044, Japan
\\\textsuperscript{7}Key Laboratory of MEMS of Ministry of Education, College of Integrated Circuits, Southeast University, Nanjing, 210096, China
\\\textsuperscript{8}State Key Laboratory of Surface Physics, Key Laboratory of Micro and Nano Photonic Structures (MOE), and Department of Physics, Fudan University, Shanghai, 200433, China
\vspace{5mm}
\\{$\dagger$} Corresponding to: wn@seu.edu.cn, jiaxj@sxu.edu.cn, liujp@shanghaitech.edu.cn, vitto.han@gmail.com, ynwang91@gmail.com 
\\{$*$} These authors contribute equally.
\vspace{5mm}
}
\end{center}
\end{minipage}
\end{center}

\setlength\parindent{13pt}
\begin{quotation}
\noindent 
\section*{Abstract}
{\textbf{Quantum Hall effects (QHE) host one-dimensional topologically-protected edge channels, which can serve as an essential ingredient in exotic quantum electronic systems. Yet the manual reconstruction of Landau-level topology, by electrostatic confinement or symmetry breaking, remains experimentally challenging. Here, we show that interfacial charge transfer in between CrOCl and large-angle twisted bilayer graphene offsets the two otherwise decoupled Dirac Landau-level ladders in each graphene layer, creating a new sequence of composite filling configurations. At charge neutrality, the composited $(+2,-2)$ state involves only the zeroth Landau levels and becomes fully insulating, with longitudinal resistance reaching the G$\Omega$ regime. By contrast, higher composite zero-filling quantum Hall states, including $(+6,-6)$ and $(+10,-10)$, retain counter-propagating helical edge channels and exhibit pronounced non-local transport, reaching up to $50\%$ of the local response. We attribute such switching-behavior to the Landau-spinor Hilbert space -- as the filling is reduced from $(+6,-6)$ to $(+2,-2)$, the orthogonal $N=\pm1$ orbital components are removed, eliminating the edge-compatible channel and gapping both bulk and boundary transport. The interaction nature of the observed gapped sates was further examined both experimentally and theoretically. Our results suggest that charge transfer provides a direct route to engineer artificial quantum Hall insulators, opening possibilities for wavefunction-selective control of helical edge modes.}}
\end{quotation}
}]
\newpage 
\clearpage

\section*{Introduction}
Helical edge states formed by counter-propagating one-dimensional modes with distinct internal quantum numbers provide a promising platform for suppressing backscattering and constructing circuits from topological quantum states \cite{bolotin2009observation,qi2011topological,lutchyn2010majorana,bernevig2006quantum,susstrunk2015observation,sanchez2017helical,nakayama2024observation,island2019spin}. Quantum Hall systems are particularly attractive in this regard, because edge modes of opposite chirality can be assembled from electron- and hole-like Landau levels and, in principle, extended to fractional states with non-Abelian excitations \cite{tsui1982two,das2005topologically,dutta2022distinguishing,lindner2012fractionalizing,lu2024fractional,clarke2013exotic,ronen2018robust,lee2017inducing,mong2014universal}. In this setting, the central challenge in an artificial quantum Hall insulator is therefore not only to create a quantum Hall bulk gap, but also to control whether its boundary remains conducting or becomes insulating. 

%Achieving such control at fixed total filling would establish the internal wavefunction structure of Landau states, rather than the macroscopic filling factor alone, as an active degree of freedom for quantum Hall engineering.

To date, several routes have been explored to superpose Landau spectra belonging to distinct Dirac sectors \cite{young2012spin,hunt2017direct,hunt2013massive}. In AA-stacked multilayer graphene, theory predicts multiple monolayer-like Dirac cones displaced in energy, whose magnetic quantization produces corresponding Landau-level ladders that can cross directly because their single-mode wavefunctions suppress inter-level coupling \cite{zhang2019landau,huang2014feature,10.1039/c5cp05013h,PhysRevB.82.165404}. Large-angle twisted bilayer graphene provides a related but experimentally accessible limit: momentum mismatch strongly suppresses interlayer hybridization, whereas the atomic-scale separation preserves substantial capacitive coupling, giving rise to two superimposed and electrostatically coupled Landau fans \cite{dong2026quantized,pezzini202030,babich2025milli,hejazi2019landau}. Dual-gated graphene electron--hole bilayers have further demonstrated counter-propagating quantum Hall edge modes \cite{huang2022valley,hoke2024uncovering,maher2013evidence,chen2020gate} and pronounced non-local transport \cite{doi:10.1126/science.1199595,jeong2024edge,gusev2012nonlocal,sui2015gate,aharon2021long}. However, a materials-level mechanism that intrinsically generates the Landau-level offset and exploits the orbital Hilbert space of the crossing states to select the resulting boundary condition has remained largely unexplored.

Here, we realize such an artificial quantum Hall insulator through interfacial charge transfer in large-angle twisted bilayer graphene. The resulting built-in electric-field gradient offsets the two otherwise approximately decoupled Dirac Landau ladders and creates a series of composite states at charge neutrality $\nu_{\mathrm{tot}}=0$, including $(+2,-2)$, $(+6,-6)$ and $(+10,-10)$. Remarkably, these states share the same total filling yet exhibit qualitatively different boundary transport. The $(+2,-2)$ configuration, involving only the zeroth Landau levels, becomes insulating in both bulk and edge, with resistance reaching the G$\Omega$ regime. In contrast, higher composite fillings retain layer-resolved counter-propagating helical edges and generate non-local responses approaching $50\%$ of the local signal. We associate this switch with the Landau-spinor wavefunctions: removal of the orthogonal $N=\pm1$ orbital components upon entering $(+2,-2)$ extinguishes the helical boundary channel. Thus, at fixed total filling, the boundary of an artificial quantum Hall insulator is selected by the Hilbert-space content of its constituent Landau states. Our results show that quantum Hall edge transport can be programmed at fixed total filling by selecting the Hilbert-space content of the constituent Landau levels through interfacial charge transfer. This wavefunction-selective control introduces a new route to engineer artificial quantum Hall insulators.

\section*{Results and Discussion}
\noindent\textbf{Charge-transfer reconstruction of decoupled Landau levels.} 
\\
Large-angle twisted bilayer graphene provides a natural platform in which two Dirac systems coexist at an atomic-scale separation while remaining largely decoupled electronically. As illustrated in Fig. 1a, the large twist angle introduces a substantial momentum mismatch between the Dirac cones of the two graphene layers, strongly suppressing direct interlayer hybridization. Each layer can therefore retain an approximately monolayer-like Dirac spectrum and, under a perpendicular magnetic field, develop an independent sequence of Landau levels (LLs). In the absence of an appreciable electrostatic asymmetry between the two layers, the corresponding Landau ladders are nearly degenerate in energy, as schematically shown in Fig. 1b. 

Figure 1c shows the dual-gated heterostructure used in this study, in which the 5-degree twisted bialyer graphene (TBLG) is placed directly on CrOCl while remaining separated from the top gate by h-BN. Detailed fabrication procedures and AFM topographic characterization of the heterostructure stacks are presented in Supplementary Figures 1-2, respectively. Charge redistribution at the graphene/CrOCl interfacial states take place across the bilayer under a proper displacement field, as described by our self-consistent electrostatic screening model in Methods (see also Supplementary Note 1 for more details). In short, charge transfer modifies their relative electrostatic energies, producing a finite relative shift of their Dirac-point energies, defined as $\Delta_{\mathrm{shift}}=E_{\mathrm{D}}^{\mathrm{bottom}}-E_{\mathrm{D}}^{\mathrm{top}}$. The resulting electronic structure can therefore be viewed as two nearly independent Dirac Landau ladders displaced with respect to one another in energy (Fig. 1d).

\begin{figure*}[ht!]
	\centering
	\includegraphics[width=0.9\linewidth]{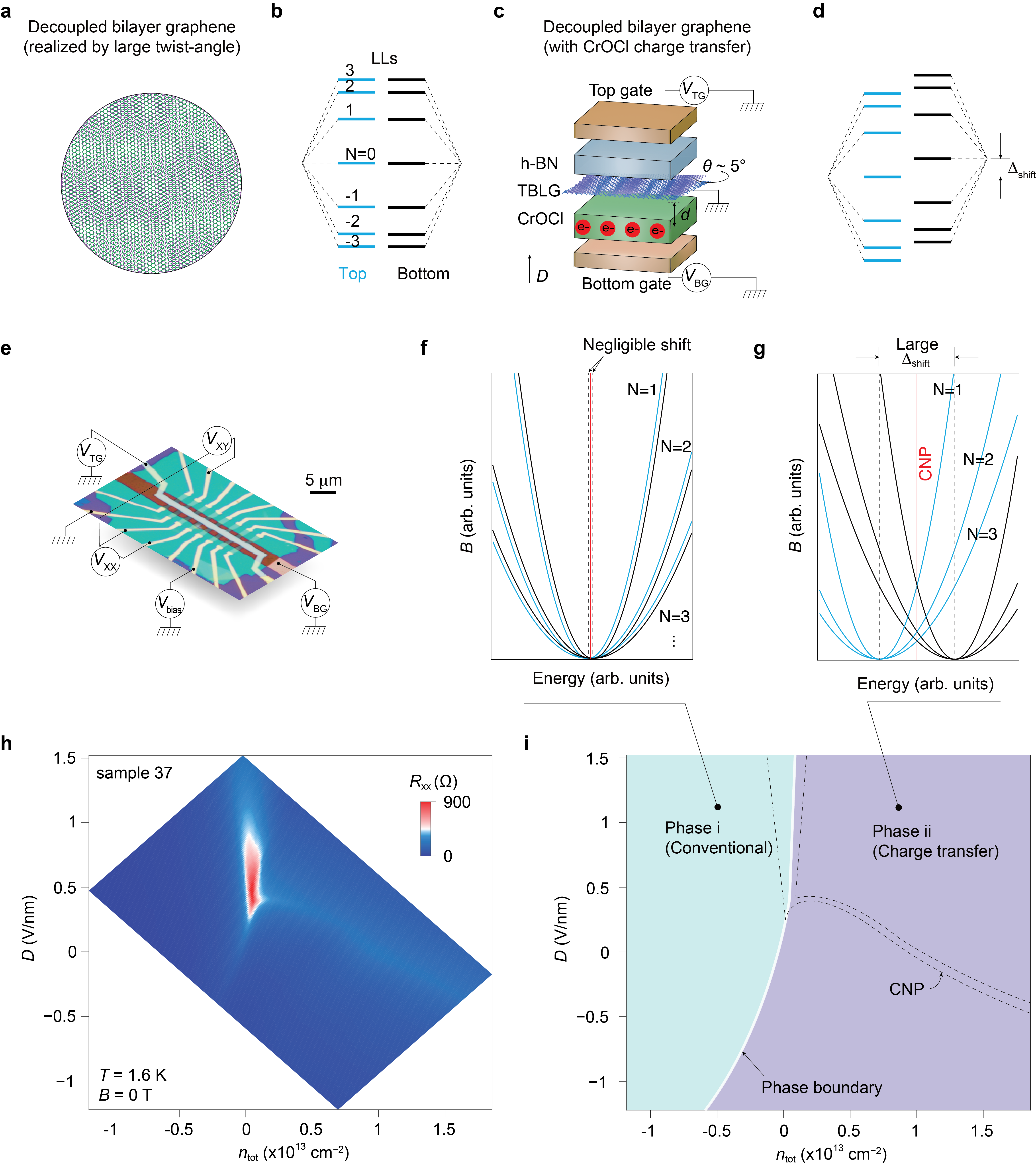}
	\caption{
		\textbf{Charge-transfer-induced reconstruction of Landau levels in large-angle twisted bilayer graphene.} \textbf{a}, Schematic of large-angle twisted bilayer graphene, in which the two graphene layers are electronically decoupled by the large momentum mismatch between their Dirac cones. \textbf{b}, Corresponding Landau-level (LL) structure of the top and bottom graphene layers, showing two nearly energy-aligned sets of LLs in the absence of appreciable interlayer energy offset. \textbf{c}, Schematic of the dual-gated large-angle twisted bilayer graphene device coupled to a CrOCl substrate. Interfacial charge transfer generates a built-in electrostatic gradient across the two graphene layers, in addition to the externally controlled displacement field $D$. \textbf{d}, Resulting LL structure, in which the two otherwise decoupled Landau ladders acquire a finite relative energy shift, $\Delta_{\mathrm{shift}}$. \textbf{e}, Optical micrograph of a representative device corresponding to the structure in \textbf{c}, with the electrical configuration used for longitudinal, transverse and non-local transport measurements. Scale bar, $5~\mu\mathrm{m}$. \textbf{f}, Schematic magnetic-field--energy spectrum for the conventional regime, where the two sets of LLs remain nearly aligned and the relative energy shift is negligible. \textbf{g}, Corresponding spectrum in the charge-transfer regime, where a pronounced $\Delta_{\mathrm{shift}}$ separates the two Landau ladders and reconstructs their crossings near the charge-neutrality point (CNP). \textbf{h}, Measured longitudinal resistance $R_{xx}$ as a function of total carrier density $n_{\mathrm{tot}}$ and displacement field $D$ at $B=0~\mathrm{T}$ and $T=1.6~\mathrm{K}$, revealing two distinct electronic regimes. \textbf{i}, Schematic phase diagram extracted from \textbf{h}, identifying the conventional phase (phase i, cyan-colored region in \textbf{i}) and the charge-transfer phase (phase ii, purple-colored region in \textbf{i}), together with the phase boundary and the CNP trajectory. The spectra in \textbf{f} and \textbf{g} correspond respectively to representative states in phases ii and ii.
	}
	\label{fig:fig1}
\end{figure*}

In a typical TBLG-CrOCl device (micrograph image is shown in Fig. e, and details of fabrications can be seen in Methods), the CrOCl layer serves as a reservoir that allows charges to be transferred in and out of its surface state, which can further form electronic crystals as reported in previous works \cite{han-graphene-crocl-arxiv21}. The Dirac points of the two layers remain nearly coincident in the conventional regime, and their LL branches are well-overlapped (Fig. 1f). On the contrary, in the charge-transfer regime, the finite $\Delta_{\mathrm{shift}}$ separates the centers of the two Landau fans along the energy axis (Fig. 1g and Supplementary Note 1). Electron-like LLs from one layer can consequently intersect hole-like LLs from the other at a series of distinct energies and magnetic fields. The system thus acquires combinations of layer-resolved quantum Hall fillings that are not obtained from a single Dirac spectrum alone. Of particular importance are composited charge-neutral configurations for which the two layers carry equal and opposite filling factors, such that
\[
\nu_{\mathrm{tot}}
=
\nu_{\mathrm{top}}
+
\nu_{\mathrm{bottom}}
=
0.
\]
This composited charge-neutrality point (CNP) thus creates a family of nominally equivalent $\nu_{\mathrm{tot}}=0$ quantum Hall states whose boundary properties nevertheless depend on the microscopic Landau wavefunctions participating in the crossing, as will be discussed in the coming sections.

Figure 1h displays the measured longitudinal resistance $R_{\mathrm{xx}}$ as a function of total carrier density $n_{\mathrm{tot}}$ and displacement field $D$ at $T = 1.6$ K and $B = 0$ T. Rather than exhibiting a single continuously evolving charge-neutrality feature as usually seen, the resistance map separates into two distinct regimes. The corresponding schematic phase diagram is indicated in Fig. 1i. At the left side of the phase boundary, phase-i represents the conventional large-angle twisted bilayer graphene regime, in which the relative energy displacement between the two Dirac systems remains small. Upon crossing the phase boundary to its right side, the device enters phase-ii, where interfacial charge transfer produces a pronounced relative energy offset between the two layers, while electronic crystals and exotic Coulomb interactions may exist as well \cite{lu-nc23}. The charge-neutrality trajectory evolves in an unusual manner within the $D-n$ landscape, as indicated by the black dashed lines in Fig. 1i. More characterizations of typical TBLG-CrOCl devices can be seen in Supplementary Figures 3-4.

 \begin{figure*}[ht!]
 	\centering
 	\includegraphics[width=0.9\linewidth]{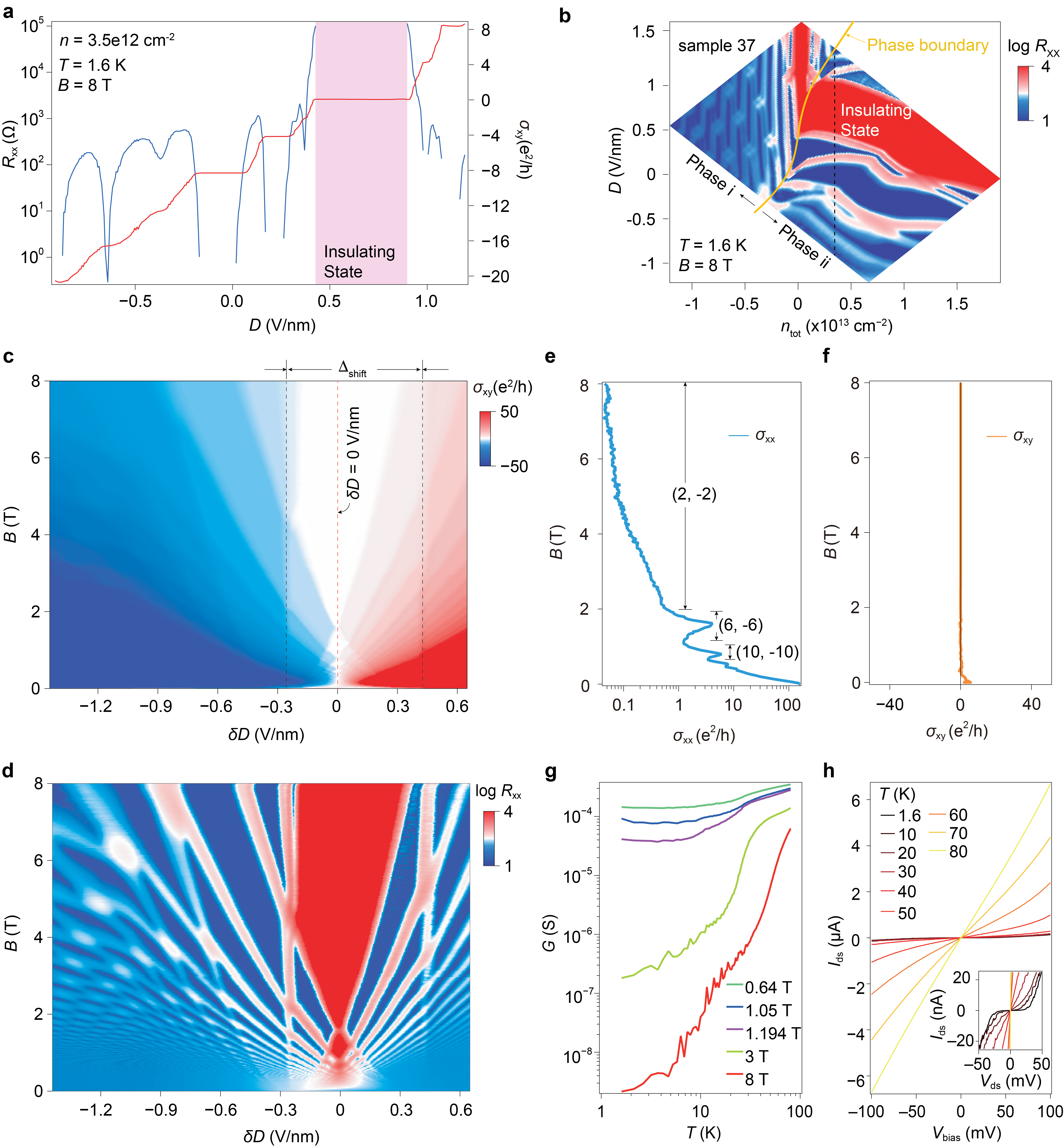}
 \caption{\textbf{Charge-transfer-induced Landau-level crossings and the insulating state at charge neutrality.}
 	\textbf{a}, Longitudinal resistance $R_{\mathrm{xx}}$ and transverse Hall response $\sigma_{xy}$ as a function of displacement field $D$ at $B=8~\mathrm{T}$ and $T=1.6~\mathrm{K}$, measured at a fixed total carrier density $n=3.5\times10^{12}~\mathrm{cm}^{-2}$. The shaded region marks the strongly insulating state.
 	\textbf{b}, Longitudinal resistance $R_{\mathrm{xx}}$ mapped in the $D$--$n_{\mathrm{tot}}$ plane at $B=8~\mathrm{T}$ and $T=1.6~\mathrm{K}$. The high-resistance region emerging in the charge-transfer phase is highlighted. The dashed line indicates the trajectory corresponding to $\textbf{a}$ and to the field-dependent measurements in the following panels.
 	\textbf{c}, Hall conductivity $\sigma_{\mathrm{xy}}$ as a function of magnetic field $B$ and relative displacement field $\delta D$, plotted over the same displacement-field range as in \textbf{a}. In the charge-transfer phase, $\delta D$ evolves approximately with the relative energy displacement of the two Dirac systems, revealing two offset families of Landau levels. The central white region corresponds to an effective total filling factor $\nu_{\mathrm{tot}}=0$. The red dashed line marks the charge-neutral trajectory, whereas the grey dashed lines indicate the characteristic energy offset $\Delta_{\mathrm{shift}}$.
 	\textbf{d}, Corresponding longitudinal-resistance map $R_{xx}(B,\delta D)$, showing two displaced and intersecting Landau fans together with the strongly insulating region centred around charge neutrality.
 	\textbf{e}, Longitudinal conductivity $\sigma_{\mathrm{xx}}$ extracted along the charge-neutral trajectory indicated by the red dashed line in \textbf{c}. As the composited fillings evolve from higher-order states, including $(+10,-10)$ and $(+6,-6)$, towards $(+2,-2)$, $\sigma_{\mathrm{xx}}$ is progressively suppressed and approaches the insulating regime.
 	\textbf{f}, Hall conductivity $\sigma_{\mathrm{xy}}$ along the same trajectory, remaining close to zero over the corresponding magnetic-field range, consistent with $\nu_{\mathrm{tot}}=0$.
 	\textbf{g}, Temperature dependence of the two-probe dc conductance at $\delta D = 0$ for selected magnetic fields. \textbf{h}, Current-voltage characteristics measured at $B=8~\mathrm{T}$ for temperatures from $1.6$ to $80~\mathrm{K}$. The inset enlarges the low-current, low-bias regime, highlighting the strongly suppressed conductance at low temperature.
 }
    \label{fig:fig2}
 \end{figure*}

\begin{figure*}[ht!]
 	\centering
 	\includegraphics[width=0.88\linewidth]{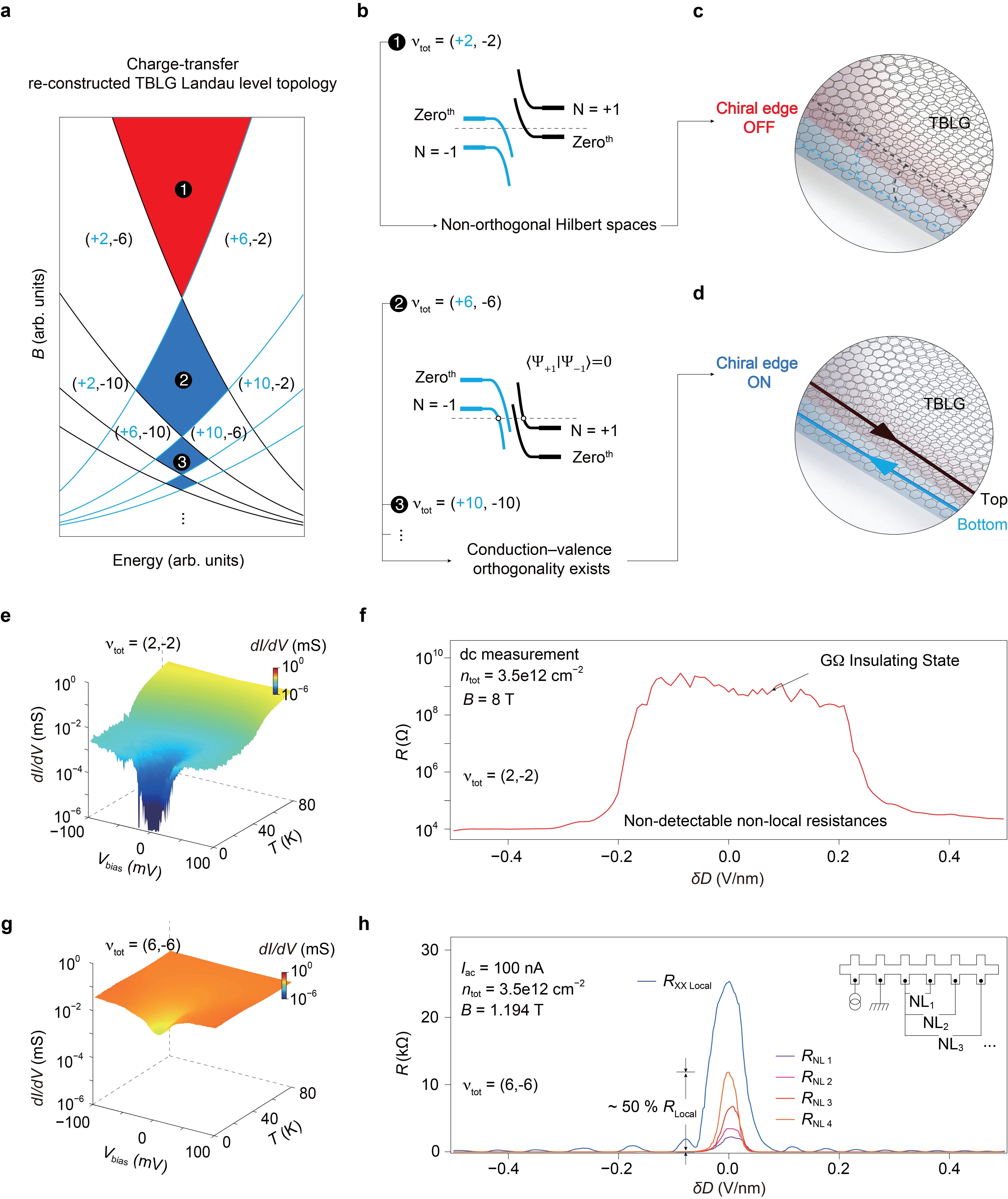}
 	\caption{\textbf{Hilbert-space-selected switching of edge transport at composite zero filling.}
	\textbf{a}, Schematic Landau-level topology reconstructed by interfacial charge transfer. The two energy-shifted Dirac Landau spectra generate a sequence of composite quantum Hall states at $\nu_{\mathrm{tot}}=0$, including $(+2,-2)$, $(+6,-6)$ and $(+10,-10)$. Numbered regions indicate representative crossings.
	\textbf{b}, Schematic Landau-levels forming the $(+2,-2)$ and $(+6,-6)$ states. For $(+2,-2)$, the crossing involves only the zeroth Landau levels and therefore lacks conduction--valence orbital orthogonality. For $(+6,-6)$, the Landau spinors are orthogonal, protecting the chiral edge. The same orthogonality extends to higher-order composite states such as $(+10,-10)$, with their cartoon illustration indicated in \textbf{c-d}. \textbf{e}, Differential conductance $dI/dV$ as a function of source--drain bias $V_{\mathrm{bias}}$ and temperature $T$ in the $(+2,-2)$ state at $\delta D=0$. \textbf{f}, dc $R_{\mathrm{xx}}$ measured across the $(+2,-2)$ composite state reaches the G$\Omega$ regime while the non-local response becomes undetectable, indicating simultaneous suppression of bulk and boundary transport. \textbf{g}, $dI/dV$ as a function of $V_{\mathrm{bias}}$ and $T$ in the $(+6,-6)$ composite state at $\delta D=0$. In contrast to \textbf{e}, a finite low-temperature conductance persists around zero bias. \textbf{h}, Local longitudinal resistance $R_{\mathrm{xx}}^{\mathrm{Local}}$ and non-local resistances $R_{\mathrm{NL}}$ across the $(+6,-6)$ state at $B=1.194$ T. A pronounced non-local response accompanies the finite local resistance at the composite $\nu_{\mathrm{tot}}=0$ state, reaching approximately $50\%$ of the local signal and evidencing transport by counter-propagating layer-resolved edge modes. The inset illustrates the non-local measurement configurations.}
 	\label{fig:fig3}
 \end{figure*}

\vspace{5mm}
\noindent\textbf{A zero-filling quantum Hall insulator from offset Landau spectra.} 
\\
Figure 2a shows the longitudinal resistance $R_{\mathrm{xx}}$ together with the Hall conductivity $\sigma_{\mathrm{xy}}$ as a function of displacement field at $B=8~\mathrm{T}$. A broad high-resistance region develops on the charge-transfer side (i.e., phase-ii) of the device, while the Hall response approaches zero over the same interval. The insulating behaviour forms an extended region inside phase-ii, shown in the $D$-$n_{\mathrm{tot}}$ map in Fig. 2b, distinct from the conventional quantum Hall states in phase-i (on the left side of phase boundary indicated by yellow solid line). Gate leakage characteristics are presented in Supplementary Figure 5.

To resolve its Landau-level origin, we track the transport response in a typical device (sample-S37) as a function of magnetic field and relative displacement field $\delta D$. Within phase-ii, $\delta D$ provides an approximately monotonic measure of the relative energetic position of the two graphene layers, and can therefore be used as an effective energy coordinate for the displaced Dirac spectra, as discussed in Supplementary Note 1. Indeed, the Hall-conductivity map in Fig. 2c reveals two families of Landau levels whose centres are separated by a finite $D_{\mathrm{shift}}$ (which is related to energy shift $\Delta_{\mathrm{shift}}$ in the chrage transfer regime, see Supplementary Note 1) of about 0.7 V/nm (slightly smaller than that of the theoretically calculated value in Supplementary Note 1). Their field evolution closely reproduces the two offset Landau ladders anticipated in Fig. 1g. In this representation, the Landau branches acquire the characteristic parabolic form in the $B$--$\delta D$ plane, and repeated intersections arise between electron-like states of one layer and hole-like states of the other.

The central region between the two sets of Landau fans is distinguished by an effective total filling factor $\nu_{\mathrm{tot}}=0$ as evidenced by the nearly vanishing Hall conductivity in Fig. 2c (white colored region). Here, $\nu_{\mathrm{tot}}=0$ correspond to a  sequence of composited zero total fillings of ($\nu_{\mathrm{bottom}}$, $\nu_{\mathrm{top}}$) =
(+2,-2), \quad (+6,-6), \quad(+10,-10), ... The corresponding longitudinal-resistance map in Fig. 2d provides a direct transport visualization of this evolution. At lower fields, the two displaced Landau spectra generate a dense network of intersecting quantum Hall states. With increasing magnetic field, the composited crossings become progressively separated until a pronounced high-resistance region develops around $\delta D=0$. Consistent transport behaviour is also observed in dc measurements of Sample-S3 (Supplementary Figure 6a-b). Tilted-field Landau fan diagrams and the temperature dependence of the Landau fan and Hall conductivity at $\theta = 63.4349^\circ$ are presented in Supplementary Figures 7-8.

This evolution is more clearly exposed by following the charge-neutral trajectory indicated by the red dashed line in Fig. 2c. The extracted longitudinal conductivity $\sigma_{xx}$ is shown in Fig. 2e. For composite states involving higher Landau levels, including approximately $(+10,-10)$ and $(+6,-6)$, a finite longitudinal conductance remains. As the magnetic field is increased and the filling is reduced towards $(+2,-2)$, $\sigma_{xx}$ decreases drastically and eventually approaches the insulating limit (with negligible conductivity). The Hall conductivity remains close to zero throughout the same evolution (Fig. 2f). The transition therefore occurs within the same macroscopic quantum Hall filling fraction $\nu_{\mathrm{tot}}=0$ (the same analysis of Sample-S62 can be found in Supplementary Fig. 9). This distinction is central to the physics discussed below -- states that are indistinguishable by their total Hall quantum number can nevertheless possess fundamentally different longitudinal and boundary transport.

Figure 2g further shows the temperature dependences of the 2-probe conductance at $\delta D=0$ for several characteristic magnetic fields. At all typical fields, the system manifests strongly suppressed conduction upon cooling, indicating insulating states. At $B=8$ T (deep in the $(+2,-2)$ zero-filling phase), the current-voltage characteristics develop an especially strongly nonlinear low-temperature response (Fig. 2h and its inset). The low-bias conductance becomes extremely small at $1.6$ K and gradually recovers as the temperature is raised, as highlighted by the enlarged low-current region in the inset. The current–voltage characteristics of the $(+6,-6)$ and $(+10,-10)$ states are presented in Supplementary Figures 10a. The corresponding thermal activation gaps are extracted from the temperature dependence of the longitudinal transport, following $I_{\mathrm{ds}} \propto \exp\left(-\frac{\Delta}{2k_{\mathrm{B}}T}\right)$, for both the $(+6,-6)$ and $(+2,-2)$ states (Supplementary Figure 10b-c).

\vspace{5mm}
\noindent\textbf{Quantum quench of helical edge modes.} 
\\
The observed distinction in conductance among different composite $\nu_{\mathrm{tot}}=0$ points to the internal Hilbert-space structure of the Landau states forming each composite configuration. To elucidate its origin, we now show that the finite conductance of the higher-order composited $\nu_\mathrm{tot}$=0 states originates predominantly from counter-propagating boundary modes, whereas their disappearance in the $(+2,-2)$ state marks a switch to a quantum Hall insulator that is insulating in both its bulk and its edge. A particularly sensitive probe of such edge-dominated transport is the non-local resistance, for which a voltage response is detected far from the current injection path only when current can propagate efficiently along the sample boundary.

A charge-transfer reconstructed Landau-level spectrum is illustrated in Fig. 3a. The energetic displacement of the two decoupled Dirac spectra produces a hierarchy of composite zero-filling states. The lowest composite state, $(+2,-2)$, is special because both the electron-like and hole-like branches at the chemical potential originate from the zeroth Landau level. By contrast, the crossings in the $(+6,-6)$ state involves the $N=+1$ and $N=-1$ Landau spinors, and higher composite fillings successively involve higher conduction- and valence-band Landau indices (Fig. 3b). For non-zero Landau index, the electron- and hole-like graphene Landau spinors possess an intrinsic conduction--valence orthogonality. In a schematic single-valley representation,
\[
\Psi_{+N}
\sim
\frac{1}{\sqrt{2}}
\begin{pmatrix}
	\phi_{|N|-1}\\
	\phi_{|N|}
\end{pmatrix},
\qquad
\Psi_{-N}
\sim
\frac{1}{\sqrt{2}}
\begin{pmatrix}
	\phi_{|N|-1}\\
	-\phi_{|N|}
\end{pmatrix},
\]
such that
\[
\langle\Psi_{+N}|\Psi_{-N}\rangle=0,
\qquad |N|\geq1.
\]
This cancellation is absent for the zeroth Landau level, whose spinor contains only a single non-vanishing orbital component. Consequently, the $N=0$ states forming $(+2,-2)$ occupy compatible orbital Hilbert spaces and can have much stronger hybridizations than the higher LL pairs when the two layer-resolved boundary modes are brought into proximity (Fig. 3c). Meanwhile, the $N=+1$ and $N=-1$ components forming $(+6,-6)$ remain orthogonal, strongly suppressing the interlayer hybridization matrix element of the edge states, thus preserving the helical edge mode transport (Fig. 3d). The charge-transfer-reconstructed Landau spectrum therefore provides a direct means of selecting helical edge transport (more details can be seen in Supplementary Note 2).

Figure 3e shows the differential conductance $dI/dV$ at $\delta D=0$ as a function of bias and temperature. A deep conductance minimum develops around zero bias and becomes progressively sharper upon cooling. The low-bias differential conductance is suppressed by several orders of magnitude at $1.6 \mathrm{K}$ compared with the high-temperature response. The dc transport measurement provides an even more striking manifestation of this state. At $B=8$ T, the longitudinal resistance $R_{\mathrm{xx}}$ rises into the G$\Omega$ range over an extended interval around the CNP, shown in Fig. 3f. Sample-S3 similarly exhibits G$\Omega$-scale resistance (Supplementary Figure 6c-d), further confirming the insulating nature of such state. Crucially, no detectable non-local resistance accompanies this enormous local resistance. 

The behaviour changes qualitatively once the chemical potential reaches the higher-order $(+6,-6)$ composite state. The differential-conductance surface in Fig. 3g retains a substantial low-bias response even at the lowest temperatures, in clear contrast to the deep insulating minimum of $(+2,-2)$ (more discussions can be seen in Supplementary Figure 11). Only a moderate zero-bias suppression rather than the many-orders-of-magnitude collapse observed for the zeroth-Landau-level state. More importantly, this residual conduction carries a pronounced spatially non-local signature. Figure 3h compares the local $R_{\mathrm{xx}}$ response with several non-local configurations across the $(+6,-6)$ state. All non-local traces develop peaks at the same composite CNP, with the largest signal reaching approximately $50\%$ of the local resistance. Such a large remote response is incompatible with ordinary diffusive bulk conduction and instead indicates that current is transported over macroscopic distances along the sample boundary. The opposite electron- and hole-like fillings of the two layers give these boundary channels opposite propagation directions, producing a helical edge configuration. Notably, dc measurements of non-local transport (Supplementary Figure 12a, b) yield the same characteristic response, further supporting the boundary-mediated transport scenario. The temperature dependence of the non-local signal (Supplementary Figure 12c) further elucidates the evolution of the edge-state contribution with temperature.

  \begin{figure}[ht!]
  \centering
 	\includegraphics[width=0.98\linewidth]{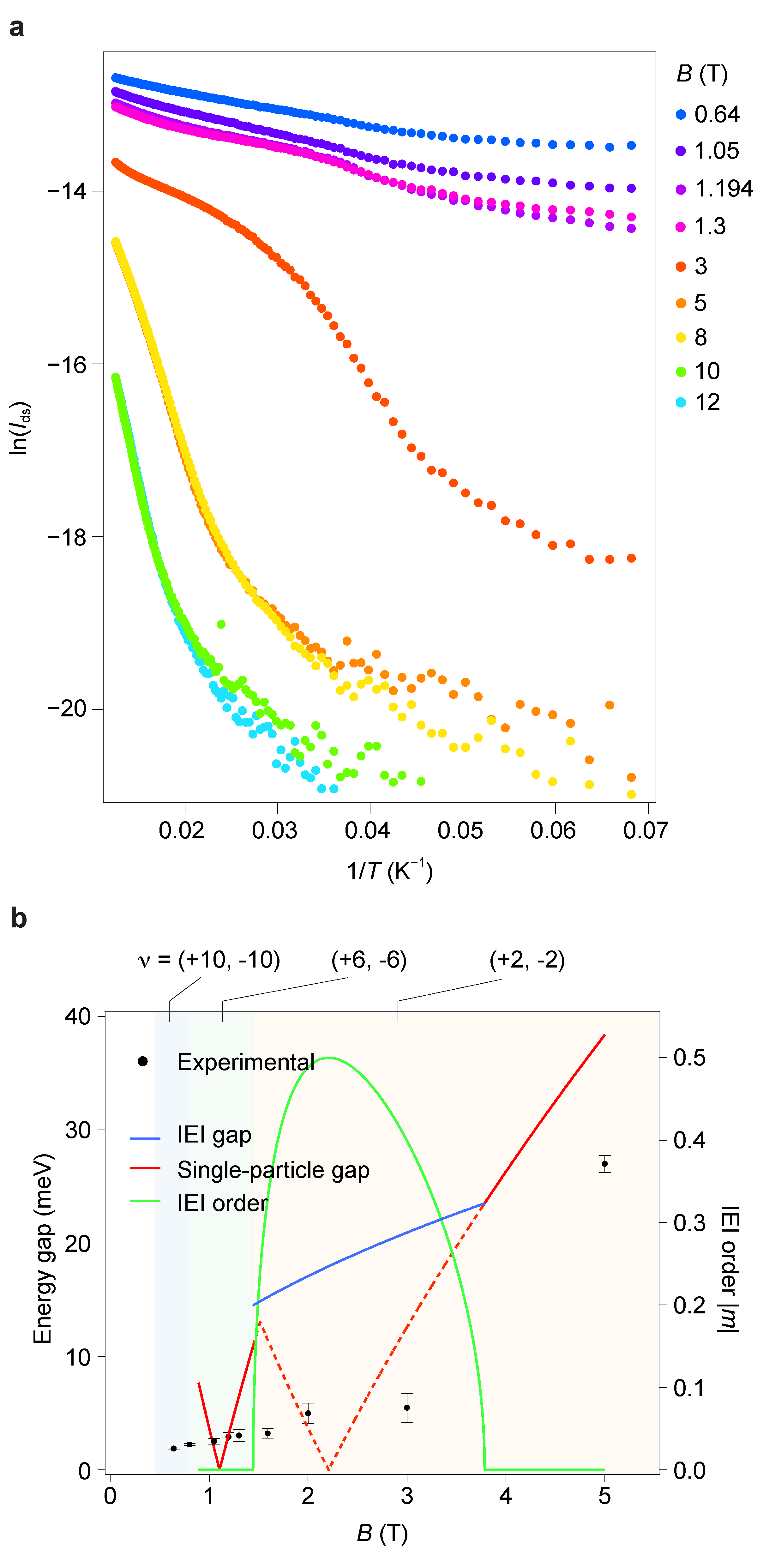}
    \caption{\textbf{Gap evolution of the artificial quantum Hall insulator.} \textbf{a}, Arrhenius plots of the source--drain current, $\ln(I_{\mathrm{ds}})$ versus $1/T$, measured at the charge-neutrality point for selected magnetic fields. The thermally activated regimes are used to extract the activation energy $\Delta E_{\mathrm{a}}$. \textbf{b}, Magnetic-field dependence of the experimentally extracted activation gap (black circles), compared with the calculated interlayer excitonic-insulator (IEI) gap (blue), single-particle Landau-level gap (red) and normalized IEI order parameter $|m|$ (green). The shaded regions indicate the successive composite quantum Hall sectors $(+10,-10)$, $(+6,-6)$ and $(+2,-2)$. Upon entering the $(+2,-2)$ sector, interlayer coherence develops and the IEI gap dominates over the bare single-particle contribution. At higher magnetic fields, the IEI order collapses and the rapidly increasing single-particle gap becomes dominant, in agreement with the experiment.
    }
 	\label{fig:fig4}
 \end{figure}

\vspace{5mm}
\noindent\textbf{On the insulating phases in composite (2,-2) quantum Hall state.} 
\\
To further study the gapped states of the artificial quantum Hall insulator at composite filling of $\nu_\mathrm{tot}=0$, we examine the energy scale associated with the insulating states. Figure 4a shows the temperature dependence of the source-drain current at the CNP for several representative magnetic fields. The activated portions of the curves become progressively steeper with increasing field, indicating a continuous enhancement of the charge-excitation gap. We extract the corresponding activation energy $\Delta E_{\mathrm{a}}$ from Arrhenius fits to the thermally activated transport, with $\ln I_{\mathrm{ds}}$ exhibiting an approximately linear dependence on $1/T$ over the selected temperature ranges.

The extracted activation energies reveal a pronounced two-stage evolution with magnetic field (black circiles indicated in Fig. 4b). At low fields, $\Delta E_{\mathrm{a}}$ remains in the few-meV range and increases only gradually as the CNP ($\delta D=0$) evolves through the successive Landau-level crossings associated with the higher-order composite quantum Hall states. A qualitatively different behaviour emerges once the system enters the lowest $(+2,-2)$ composite sector. Beyond approximately 4 T, the activation gap increases much more rapidly,reaching values above $30 \mathrm{meV}$. The magnitude and field dependence of this enhancement are distinct from the smooth evolution observed across the lower-field Landau-level crossings, indicating that the strongly insulating $(+2,-2)$ state cannot be described simply as a continuation of the same gap-opening mechanism.

To understand the two-stage evolution, we consider the charge-transfer-reconstructed Landau spectrum together with interlayer Coulomb interactions. Upon entering the $(+2,-2)$ composite quantum Hall sector, the $n=-1$ and $n=+1$ Landau levels become the active pair closest to the chemical potential. Projecting the Coulomb interaction onto these levels yields an interlayer Fock exchange that can stabilize an interlayer excitonic insulating (IEI) state when the exchange energy exceeds their effective detuning (Supplementary Note 2). The IEI order is strongest near the Landau-level crossing, opening an interaction-driven gap where the bare single-particle gap is small. With increasing magnetic field, however, the single-particle level separation grows faster than the interlayer exchange, causing the IEI order to collapse and the single-particle quantum Hall gap to become dominant. Details of the calculation are provided in Supplementary Note 2.

This simulated crossover agrees qualitatively well with the experimental activation-gap measurements in Fig. 4b. Just after the system evolves from $(+6,-6)$ into $(+2,-2)$, the insulating gap is dominated by interlayer correlations. At higher fields, the calculated IEI order vanishes and the gap crosses over to the more rapidly increasing single-particle contribution, reproducing the experimentally observed two-step enhancement from a few meV to several tens of meV. The $(+2,-2)$ artificial quantum Hall insulator therefore evolves from an interaction-driven interlayer excitonic state into a single-particle insulating state as the magnetic field is increased.

\vspace{5mm}
To conclude, we have devised an interfacial charge transfer route to construct an artificial quantum Hall states from two electrostatically decoupled whilst energetically displaced Dirac LLs systems. The resulting zero-filling composite states, including $(+10,-10)$, $(+6,-6)$ and $(+2,-2)$, share the same total filling $\nu_{\mathrm{tot}}=0$ but exhibit drastically different boundary transport. As a result, higher-order states retain layer-resolved counter-propagating edge modes and pronounced non-local conduction, whereas the lowest $(+2,-2)$ state becomes insulating in both bulk and boundary. This contrast identifies the Landau-level Hilbert space as the key variable selecting the edge condition. Within the $(+2,-2)$ sector, this insulating gap are calculated to be evolving from an interaction-driven interlayer excitonic regime to a single-particle Landau-level gap at higher magnetic fields, which echos with the experimental observations. Our results establish charge transfer as a means to engineer both the edge connectivity and many-body character of artificial quantum Hall insulators, offering a route towards wavefunction-selective control of topological boundary transport.

 \clearpage

\section*{Methods}
\vspace{3mm}
\noindent\textbf{Fabrication of devices.} 
Van der Waals heterostructures consisting of the h-BN/twisted bilayer graphene/CrOCl sandwich were fabricated from mechanically exfoliated bulk crystals. Suitable flakes were identified by optical microscopy and atomic force microscopy (AFM). A monolayer graphene flake was cut into two pieces with an AFM tip. The heterostructure was assembled by a dry-transfer technique using a polycarbonate (PC) film. An h-BN flake, the first graphene layer, the second graphene layer, rotated by $5^{\circ}$ relative to the first layer, and a CrOCl flake were sequentially picked up before the completed stack was released onto a pre-patterned Au gate on a $Si/SiO_{\mathrm{2}}$ substrate. Hall-bar devices were defined by reactive ion etching. Electron-beam lithography was carried out using a Zeiss Sigma 360 scanning electron microscope equipped with a Raith Elphy Quantum pattern generator. One-dimensional edge contacts and gate electrodes were fabricated by electron-beam evaporation with Cr/Au thicknesses of 5/50 nm and 5/30 nm, respectively.

\vspace{3mm}
\noindent\textbf{Electrical measurements.} 
Room-temperature electrical characterization was carried out using a probe station (Cascade Microtech EPS150) equipped with an Agilent B1500A Semiconductor Device Parameter Analyzer. Low-temperature transport measurements under high magnetic fields were performed in an Oxford Teslatron cryostat using both four-probe and two-probe configurations. For AC measurements, a low-frequency excitation current of 100 nA was applied, and the longitudinal, Hall, and non-local voltages were measured using Stanford Research Systems SR860 lock-in amplifiers. For DC measurements, a Keithley 2636B source meter was used to apply a bias voltage of 5 mV while simultaneously measuring the corresponding voltages. Gate voltages were supplied by a Keithley 2400 source meter.

\subsection*{Theoretical modelings.}

The charge-transfer-induced band reconstruction of the CrOCl-TBG heterostructure was calculated using a self-consistent electrostatic screening model. The non-interacting Hamiltonian was written as
\begin{equation}
H^{0}=H_{\mathrm{TBG}}^{0}+H_{\mathrm{CrOCl}}^{0},
\end{equation}
where $H_{\mathrm{TBG}}^{0}$ is the Bistritzer-MacDonald continuum Hamiltonian for twisted bilayer graphene \cite{macdonald-pnas11,koshino-prx18}. The low-energy CrOCl surface state was approximated by an isotropic two-dimensional parabolic band,
\begin{equation}
H_{\mathrm{CrOCl}}^{0}(\mathbf{k})
=
\frac{\hbar^{2}k^{2}}{2m^{*}}+E_{\mathrm{CBM}},
\end{equation}
with $m^{*}=1.3m_{0}$ and
$E_{\mathrm{CBM}}=-0.13~\mathrm{eV}$ \cite{lu-nc23,han-nn2022}.
Direct hopping between TBG and CrOCl was neglected owing to the large interlayer distance and lattice mismatch.

For an external displacement field $D$, the bare electrostatic potential of graphene layer $\ell$ was taken as
\begin{equation}
U_{\ell}
=
e\frac{D}{\epsilon_{r}}
\left[d_{s}+(\ell-1)d_{\mathrm{g}}\right],
\qquad \ell=1,2,
\end{equation}
where $d_{s}=4.5~\text{\AA}$ is the CrOCl--graphene separation and $d_{\mathrm{g}}=3.35~\text{\AA}$ is the graphene interlayer distance. The layer-resolved charge densities obtained from the continuum Hamiltonian were used as source terms in the Poisson equation to update the screened electrostatic potentials. The leading exchange correction
to the CrOCl surface state was included through
\begin{equation}
E_{\mathrm{CBM}}^{*}
=
E_{\mathrm{CBM}}
-
\frac{e^{2}\sqrt{\pi n_{\mathrm{CrOCl}}}}
{4\pi\epsilon_{0}\epsilon_{r}^{\mathrm{CrOCl}}}.
\end{equation}
Remote-band Coulomb corrections to the TBG parameters were incorporated during the self-consistent screening procedure using a perturbative renormalization-group treatment \cite{kang-rg-prl20,guo-prb24}. The screened potentials, layer-resolved carrier densities, CrOCl Fock shift, Fermi energy and renormalized TBG parameters were iterated until convergence.

Self-consistent calculations were performed on a $36\times36$ $k$-point mesh of the moir\'e Brillouin zone as a function of $D$ at fixed total carrier density $n_{\mathrm{tot}}=3.5\times10^{12}~\mathrm{cm}^{-2}$. The Dirac-point energy $E_{D}^{(\ell)}$ of each graphene layer was identified from the condition of vanishing layer-resolved excess carrier density, and the relative charge-transfer-induced energy offset was defined as
\begin{equation}
\Delta=E_{D}^{(1)}-E_{D}^{(2)}.
\end{equation}

For the large twist angle considered here, the moir\'e potential was treated perturbatively in calculating the Landau spectrum. The effective Dirac velocity was obtained from
\begin{equation}
\frac{v_{F}^{*}}{v_{F}^{0}}
=
\frac{1-3(\alpha_{1}^{0})^{2}}
{1+3\left[(\alpha_{0}^{0})^{2}+(\alpha_{1}^{0})^{2}\right]},
\qquad
\alpha_{j}^{0}
=
\frac{w_{j}}{\hbar v_{F}^{0}k_{\theta}},
\qquad
k_{\theta}=2K_{D}\sin\frac{\theta}{2},
\end{equation}
which gives
$\hbar v_{F}^{*}\simeq4.64~\mathrm{eV\,\text{\AA}}$ for
$\theta=5^{\circ}$, $\hbar v_{F}^{0}=5.25~\mathrm{eV\,\text{\AA}}$,
$w_{0}=79.7~\mathrm{meV}$ and $w_{1}=97.5~\mathrm{meV}$
\cite{TBGI-Bernervig}. The layer-resolved graphene Landau levels were
then calculated as
\begin{equation}
\varepsilon_{n\ell}(B)
=
E_{D}^{(\ell)}
+
\operatorname{sgn}(n)v_{F}^{*}
\sqrt{2e\hbar B|n|},
\qquad
n=0,\pm1,\pm2,\ldots .
\end{equation}
The resulting two displaced Dirac Landau-level sequences constitute the Landau spectrum used in the main-text analysis. Further details of the continuum model and self-consistent screening procedure are provided in Supplementary Note 1.

Interlayer excitonic interactions near the crossing of the $N=-1$ and $N=+1$ graphene Landau levels were further treated by projecting the long-range Coulomb interaction onto the $N=\pm 1$ Landau-level wavefunctions based on Hartree-Fock approximations. Details of the models, numerical parameters and self-consistent procedures are provided in Supplementary Notes 2.

\clearpage

\section*{\label{sec:level1}Data Availability}

The data that support the findings of this study will be available with a Zenodo doi link when published.

\section*{\label{sec:level2}Code Availability}

The code that support the findings of this study are available upon reasonable request to the corresponding authors.

\section*{\label{sec:level3}Acknowledgements}
This work is supported by the National Key R$\&$D Program of China (No. 2022YFA1203903) and the National Natural Science Foundation of China (NSFC) (Grant Nos. 92265203, 12104462, 11974357 and 12304237). Z.V.H. acknowledges the support of the Fund for Shanxi “1331 Project” Key Subjects Construction, and the Innovation Program for Quantum Science and Technology (grant no. 2021ZD0302003). Kenji Watanabe and Takashi Taniguchi acknowledge support from the JSPS KAKENHI (Grant Numbers 21H05233 and 23H02052) , the CREST (JPMJCR24A5), JST and World Premier International Research Center Initiative (WPI), MEXT, Japan. Neng Wan acknowledges the open project of key Laboratory of Artificial Structures and Quantum Control (Ministry of Education), Shanghai Jiao Tong University, and the Postgraduate Research $\&$ Practice Innovation program of Jiangsu province KYCX22$\_$0228.

\section*{Author Contributions}
Y.W., Z.V.H., J.L., X.J., and N.W. conceived the experiment and supervised the overall project. N.R., Y.W., K.Y., and Z.X. performed the device fabrications and electrical measurements; Y.S. and J.L. performed the theoretical modelings. Z.V.H., Y.W., Y.S., H.W., X.J., and J.L. analyzed the experimental data. K.W., T.T., and N.W. provided the high quality h-BN crystal in this study. The manuscript was written by Z.V.H., Y.W., Y.S. and J.L. with discussions and inputs from all authors.

\section*{Competing Interests}
The authors declare no competing interests.

\end{document}